\documentclass[prl,twocolumn,superscriptaddress,nofootinbib]{revtex4-1}

\pdfoutput=1
\usepackage{amssymb,amsmath,amsthm,graphicx}
\usepackage{subfigure}
\usepackage{graphics, color}
\usepackage{latexsym}
\usepackage{bm}
\usepackage{epsfig}

\def \dd {\mathrm{d}}

\begin{document}
\title{Stringy corrections to the entropy of electrically charged\\
supersymmetric black holes with $\mathrm{AdS}_5\times S^5$ asymptotics.}
\author{Jo\~ao~F.~Melo}
\email{jfm54@cam.ac.uk}
\affiliation{DAMTP, Centre for Mathematical Sciences, University of Cambridge, Wilberforce Road, Cambridge CB3 0WA, UK}
\author{Jorge~E.~Santos}
\email{jss55@cam.ac.uk}
\affiliation{DAMTP, Centre for Mathematical Sciences, University of Cambridge, Wilberforce Road, Cambridge CB3 0WA, UK}
\affiliation{Institute for Advanced Study, Princeton, NJ 08540, USA}
\begin{abstract}\noindent{We study the leading $\alpha^\prime$ corrections to the entropy of certain black holes with AdS$_5\times S^5$ asymptotics. We find that, in the supersymmetric limit, the entropy does not receive $\alpha^\prime$ corrections. This result strengthens recent calculations that match the index of $\mathcal{N}=4$ Super-Yang-Mills with the corresponding partition function in the supersymmetric limit. In the small temperature regime, we find that the entropy corrections are concordant with the weak gravity conjecture.
}
\end{abstract}
\maketitle

\noindent\emph{\bf Introduction.} Quantum gravity remains a largely unexplored frontier. However, due to the seminal work in black hole thermodynamics in the seventies \cite{Christodoulou:1970wf,Hawking:1971tu,carter,Bekenstein:1972tm,Christodoulou:1972kt,Bekenstein:1973ur,Bardeen:1973gs,Hawking:1974rv,Hawking:1974sw} we know that, whatever the ultimate unifying theory is, it should reproduce the Hawking effect and give a microscopic derivation of the Bekenstein-Hawking black hole entropy in the appropriate semi-classical limit. To date, string theory appears to be the only candidate for a quantum theory of gravity that explains both of these effects in an ambiguity free manner at a microscopic level \cite{Strominger:1996sh,Breckenridge:1996sn,Callan:1996dv,Horowitz:1996fn}. In particular, the seminal work of \cite{Strominger:1996sh} provided a beautiful matching between the Bekenstein-Hawking entropy of certain five-dimensional supersymmetric black holes with asymptotically flat boundary conditions and the counting of specific supersymmetric states. Since then, a number of generalisations of this work have been accomplished for black holes with more complex topologies (see \emph{e.g.} \cite{Cyrier:2004hj}).

However, this matching has only been accomplished for black holes with asymptotically flat boundary conditions. One might wonder how to extend these results to asymptotically anti-de Sitter (AdS) spacetimes, for which we have the so-called AdS/CFT correspondence \cite{Maldacena:1997re,Gubser:1998bc,Witten:1998qj,Aharony:1999ti}. In its original form, the AdS/CFT correspondence relates four-dimensional $\mathcal{N}=4$ Super-Yang-Mills (SYM) with gauge group $SU(N)$ and 't Hooft coupling $\lambda$, to type IIB superstring theory with string coupling $g_s$, string length $\ell_s\equiv \sqrt{\alpha^\prime}$ on $\mathrm{AdS}_5\times S^5$ with radius $L$ and $N$ units of $F_{(5)}$ flux through the $S^5$. The field theory is thought to live at the conformal boundary of $\mathrm{AdS}_5$, and for this reason the correspondence is said to be holographic in nature. The string theory side is often referred to as the `bulk' and the field theory side as the `boundary'.

The parameters on each side of the AdS/CFT correspondence are related via
\begin{equation}
\frac{\lambda}{N}=2\pi g_{s}\qquad\text{and}\qquad 2\lambda = \frac{L^4}{\ell_s^4}\,.
\end{equation}

However, it remains a challenge to understand string theory for generic values of $g_{s}$, so one usually takes $N\to+\infty$, at fixed $\lambda$, so that $g_s\to0$. Under these assumptions, the bulk theory reduces to a classical theory of strings. To simplify matters further, we can also take $\lambda$ to be large, but not necessarily infinite. On the field theory side, we are thus looking at strong coupling effects, and on the gravity side we have a supergravity theory. Corrections to the strict $\lambda\to+\infty$ limit appear in the bulk as higher derivative terms which account for finite size string corrections.

The problem of reproducing the entropy of certain black hole solutions in global $\mathrm{AdS}_5$ on the string theory side is now mapped into a counting problem of certain states on the field theory. Because we are interested in global $\mathrm{AdS}_5$, the field theory is thought to live on $\mathbb{R}_t \times S^3$. The holographic description of electrically-charged supersymmetric black holes with $\mathrm{AdS}\times S^5$ asymptotics is in terms of states of the dual $\mathcal{N}=4$ SYM that preserve only one of the available sixteen supercharges. Such states should be counted (with sign) by the superconformal index. However, early attempts to compute this index gave an order one result \cite{Kinney:2005ej}, whereas the entropy of $\mathrm{AdS}_5$ black holes scales with $N^2$. It was not until recently that this long-standing problem was partially solved. In particular, \cite{Hosseini:2017mds,Cabo-Bizet:2018ehj,Choi:2018hmj,Benini:2018ywd,Honda:2019cio,ArabiArdehali:2019tdm,Zaffaroni:2019dhb,Kim:2019yrz,Cabo-Bizet:2019osg,Lezcano:2019pae,Lanir:2019abx,Cabo-Bizet:2019eaf,Cabo-Bizet:2020nkr,Murthy:2020rbd,Agarwal:2020zwm,Benini:2020gjh,Gadde:2020yah} have argued that, upon using complex chemical potentials, the cancellations between fermionic and bosonic degrees of freedom observed in \cite{Kinney:2005ej} can be avoided. This leads to an index of order $N^2$, whose associated entropy matches those of known supersymmetric black holes \cite{Gutowski:2004ez,Gutowski:2004yv,Kunduri:2006ek}. This body of work thus provides overwhelming evidence that whether we compute the entropy via the index or via a more standard calculation using the partition function of $\mathcal{N}=4$ SYM, the results should agree with each other. It should be noted that this latter quantity can only be computed via an indirect bulk calculation using the Bekenstein-Hawking entropy.


The matching between the partition function calculation and index, leads to a number of fascinating predictions. In particular, since the index cannot exhibit a dependence on continuous parameters\footnote{Except perhaps when wall-crossing is observed, see \emph{e.g.} \cite{Bates:2003vx,Dabholkar:2012nd}.}, we expect the counting on the field theory side to not depend on the 't Hooft coupling $\lambda$. On the bulk side of the story, because we are computing directly a partition function, this is not an obvious fact given we know that the classical equations of motion of type IIB supergravity do admit corrections in $\alpha^\prime$, due to finite size stringy effects. These, in the small $\alpha^\prime$ limit, appear as higher-derivative corrections to the equations of motion of type IIB supergravity. The first non-trivial corrections for supergravity configurations that only involve the metric $g$ and five-form $F_{(5)}$ were worked out in \cite{Paulos:2008tn}\footnote{We would like to note, however, that \cite{Paulos:2008tn} has a number of typos in their section 4, which summarises their results.}, following the seminal results of \cite{Green:2003an}.

\noindent\emph{\bf The black holes.} We focus on black hole solutions of five-dimensional minimal gauged supergravity, whose action comprises a five-dimensional metric $g$ and a field strength $F=\mathrm{d}A$ and reads
\begin{multline}
S_{5D}=\frac{1}{16 \pi G_5}\int_{\mathcal{M}} \mathrm{d}^5 x\sqrt{-g}\Bigg(R +\frac{12}{L^2}
\\
-\frac{1}{4}F_{ab}F^{ab}+\frac{1}{12\sqrt{3}}\varepsilon^{abcde}F_{ab}F_{cd}A_e\Bigg)\,.
\label{eq:lower}
\end{multline}
Known black hole solutions in this theory carry one electric charge $Q$, and two angular momenta $J_1$, $J_2$. For simplicity, we focus on the case where $J_1=J_2=J$. The equations of motion derived from Eq.~(\ref{eq:lower}) read
\begin{subequations}
\begin{align}
& R_{ab}-\frac{g_{ab}}{2} R-\frac{6}{L^2} g_{ab} = \frac{1}{2}\left(F_{a}^{\phantom{a}c}F_{bc}-\frac{g_{ab}}{4}F^{cd}F_{cd}\right)\,,
\\
& \nabla_a F^{ab} = \frac{1}{4\sqrt{3}}\varepsilon^{bcdef}F_{cd}F_{ef}\,.
\end{align}
\end{subequations}

We are interested in the $\alpha^\prime$ corrections to the entropy of the black holes constructed in \cite{Cvetic:2004hs}, which read
\begin{subequations}
\begin{align}
&\mathrm{d}s^2_{5D} = -\frac{f}{h}\mathrm{d}t^2+\frac{\mathrm{d}r^2}{f}+\frac{r^2}{4}(\sigma_1^2+\sigma_2^2)+\frac{r^2}{4}h\left(\sigma_3-W\, \mathrm{d}t\right)^2\,,
\label{eq:5Dg}
\\
& A = \frac{\sqrt{3}\tilde{Q}}{r^2}\left(\mathrm{d}t-\frac{\tilde{J}}{2}\sigma_3 \right)\,,
\end{align}
\label{eq:5D}%
\end{subequations}
where $\sigma_1, \sigma_2, \sigma_3$ are the usual left-invariant 1-forms of $S^3$
\begin{subequations}
\begin{align}
&\sigma_1 =-\sin\psi~\dd\theta+\cos\psi\sin\theta~\dd\phi \,,
\label{eq:sigma1}
\\
&\sigma_2=\cos\psi~\dd\theta+\sin\psi\sin\theta~\dd\phi\,,
\label{eq:sigma2}
\\
&\sigma_3=\dd\psi+\cos\theta~\dd\phi\,,
\end{align}
\label{eq:sigma3}%
\end{subequations}
and
\begin{subequations}
\begin{align}
    &f = \frac{r^2}{L^2}+1-\frac{2 \tilde{M}}{r^2}\left(1-\chi\right)+\frac{\tilde{Q}^2}{r^4}\left(1-\frac{\tilde{J}^2}{L^2}+\frac{2\tilde{M} L^2 \chi}{\tilde{Q}^2}\right)\,,
    \\
    & W = \frac{2\tilde{J}}{r^2 h}\left(\frac{2 \tilde{M}+\tilde{Q}}{r^2}-\frac{\tilde{Q}^2}{r^4}\right)\,,
    \\
    & h = 1-\frac{\tilde{J}^2 \tilde{Q}^2}{r^6}+\frac{2 \tilde{J}^2 (\tilde{M}+\tilde{Q})}{r^4}\,,
\end{align}
\end{subequations}
where $L^2\chi \equiv \tilde{J}^2(1+\tilde{Q}/\tilde{M})$. The constants $\tilde{M}$, $\tilde{Q}$ and $\tilde{J}$ parametrise the energy $M$, electric charge $Q$ and angular momentum $J$ as
\begin{subequations}
\begin{align}
&M = \frac{3 \tilde{M}\pi}{4 G_5}\left(1+\frac{\chi}{3}\right)\,,
\\
& J = \frac{\tilde{J} \pi}{4 G_5}(2\tilde{M}+\tilde{Q})\,,
\\
& Q = \frac{\sqrt{3} L \pi \tilde{Q}}{4 G_5}\,.
\end{align}
\end{subequations}

The black hole event horizon is the null hypersurface $r=r_+$, with $r_+$ being the largest real positive root of $f(r)$. The associated Hawking temperature $T$, entropy $S$, chemical potential $\mu$ and angular velocity $\Omega$ can be found in \cite{Cvetic:2004hs}. It is then a simple exercise to check that all thermodynamic quantities satisfy the first law of black hole mechanics
\begin{equation}
\mathrm{d} E = T\,\mathrm{d}S+\mu\,\mathrm{d}Q+\Omega\,\mathrm{d}J\,.
\end{equation}
The Gibbs free energy is then constructed in the usual manner via $G = E-T\,S -\mu\,Q-\Omega\,J$. One can show that $G/T$ agrees with the Euclidean on-shell action (\ref{eq:lower}) up to the usual Gibbons-Hawking-York \cite{York:1972sj,Gibbons:1976ue} term and boundary counterterms \cite{Balasubramanian:1999re,deHaro:2000vlm}.

Finally, with our normalizations for $F$, the BPS condition is given by\footnote{To avoid cluttering in the notation, from here onward we take $Q\geq 0$ and $J\geq0$.}
\begin{equation}
    \Delta \equiv M-\frac{2}{L}J-\frac{\sqrt{3}}{L}Q\geq0\,.
    \label{eq:BPS}
\end{equation}
The saturation of the BPS condition occurs only for supersymmetric solutions. Similar BPS bound have been shown not to receive $\alpha^\prime$ corrections even for asymptotically flat black holes \cite{Loges:2020trf}. The AdS BPS condition (\ref{eq:BPS}), together with the first law, implies $T=0$, $\Omega = 2/L$ and $\mu = \sqrt{3}/L$, which in turn yield
\begin{subequations}
\begin{align}
&\tilde{Q}=\tilde{Q}_{\mathrm{BPS}} \equiv r_+^2 \left(1+\frac{r_+^2}{2 L^2}\right)\,,
\\
&\tilde{J}=\tilde{J}_{\mathrm{BPS}}\equiv \frac{L r_+^2}{r_+^2+2 L^2}\,.
\end{align}%
\end{subequations}
Note that even though the solutions (\ref{eq:5D}) appear to depend on three parameters ($\tilde{M},\tilde{Q},\tilde{J}$), the BPS condition reduces this family to a one-parameter family, despite the fact that extremal black holes form a two-parameter family of solutions. We remark that \cite{Markeviciute:2016ivy,Markeviciute:2018cqs} provided strong numerical evidence for the existence of a new two-parameter family of supersymmetric black holes, whose role in this story remains to be understood. One can also show that demanding the absence of naked singularities in (\ref{eq:5Dg}) implies that $L> \tilde{J}$ \footnote{These is not the only restrictions on the three-dimensional moduli space of black hole solutions $\{\tilde{J}, \tilde{Q},r_+\}$ that bulk regularity demands, but it is the only one we will need to show that $\delta S>0$.}.

Since the $\alpha^\prime$ corrections are only know in type IIB supergravity, we uplift the solutions (\ref{eq:5D}) to ten dimensions. Using the results of \cite{Chamblin:1999tk,Cvetic:1999xp,Cvetic:2000nc}, one can show that Eq.~(\ref{eq:5D}) oxidises to the following solution of type IIB supergravity:
\begin{subequations}
\begin{align}
    &\mathrm{d}s^2 = \mathrm{d}s^2_{5D}+L^2\left[ \left(\mathrm{d}\Psi+\mathbb{A}-\frac{A}{\sqrt{3}L}\right)^2+\mathrm{d}\mathbb{CP}^2\right]
    \label{eq:metric10D}
    \\
    & G_{(5)} = \frac{r^3}{2 L}\mathrm{d}t\wedge \mathrm{d}r\wedge \sigma_1 \wedge \sigma_2 \wedge \sigma_3+\frac{L^3}{2\sqrt{3}}\; \mathbb{J}\wedge \star_5 F\,,
    \\
    & F_{(5)}=G_{(5)}+\star_{10} G_{(5)}
\end{align}
\label{eqs:uppper}%
\end{subequations}
where $\star_5$ is the five-dimensional Hodge dual obtained using the line element (\ref{eq:5Dg}), $\star_{10}$ is the Hodge dual obtained using the ten-dimensional line element (\ref{eq:metric10D}), $\mathrm{d}\mathbb{CP}^2$ is the standard Fubini-Study metric on $\mathbb{CP}^2$ and $\mathbb{J}=\mathrm{d}\mathbb{A}$ is its associated K\"ahler form.
\\
\noindent\emph{\bf Evaluating the corrections.} 
The action\footnote{As usual, we use this term with a certain abuse of notation, because the five-form $F_{(5)}$ is only made self-dual at the level of the equations of motion. After the inclusion of the correction term proportional to $\gamma$, the self-duality condition is accordingly changed.} with the leading order $\alpha'$ correction is \cite{Paulos:2008tn}:
\begin{equation}
    S_{IIB}=\frac{1}{16\pi G_{10}}\int_{\mathcal{M}_{10}}\mathrm{d}^{10}x \sqrt{-g}\left(R-\frac{1}{4\times 5!}F_{(5)}^2+\gamma\, \mathcal{W}\right)
    \label{eq:IIB}
\end{equation}
where $\mathcal{W}$ is given by
\begin{equation}
    \mathcal{W}\equiv \frac{1}{86016}\sum_{i=1}^{20} n_i M_i
\end{equation}
with all twenty monomials given in table~\ref{FinalRes} and\footnote{Note that after computing $\mathcal{T}$ with this expression, one still needs to antisymmetrise over the first three indices and the last three indices and then symmetrise for their exchange, before plugging into the monomials.}
\begin{multline}
    \mathcal T_{a b c d e f}=i \nabla_{a} F_{b c d e f}+
    \\
    \frac 1{16}\left (F_{a b c m n}F_{d e f}^{\ \ \ m n}-3 F_{a b f m n}F_{d e c}^{\ \ \ m n}\right)\,.
\end{multline}
Finally, we also have
\begin{equation}
\gamma = \frac{{\alpha^\prime}^3}{16}\frac{\pi ^3}{8} \zeta (3)\,.
\end{equation}

\begin{table}
\begin{tabular}{|c|c|}
\hline $n_i$  & $M_i$ \\ \hline
-43008 & $ C_{a b c d} C_{a b e f} C_{c e g h} C_{d g f h} $ \\ \hline
86016 & $  C_{a b c d} C_{a e c f} C_{b g e h} C_{d g f h}$ \\ \hline
129024 & $ C_{a b c d} C_{a e f g} C_{b f h i} \mathcal T_{c d e g h i}$ \\ \hline
30240 & $ C_{a b c d} C_{a b c e} \mathcal T_{d f g h i j} \mathcal T_{e f h g i j} $ \\ \hline
7392 & $ C_{a b c d} C_{a b e f} \mathcal T_{c d g h i j} \mathcal T_{e f g h i j}$ \\ \hline
-4032 & $ C_{a b c d} C_{a e c f} \mathcal T_{b e g h i j} \mathcal T_{d f g h i j}$ \\ \hline
-4032 & $ C_{a b c d} C_{a e c f} \mathcal T_{b g h d i j} \mathcal T_{e g h f i j}$ \\ \hline
-118272 & $ C_{a b c d} C_{a e f g} \mathcal T_{b c e h i j} \mathcal T_{d f h g i j}$ \\ \hline
-26880 & $ C_{a b c d} C_{a e f g} \mathcal T_{b c e h i j} \mathcal T_{d h i f g j}$ \\ \hline
112896 & $ C_{a b c d} C_{a e f g} \mathcal T_{b c f h i j} \mathcal T_{d e h g i j}$ \\ \hline
-96768 & $ C_{a b c d} C_{a e f g} \mathcal T_{b c h e i j} \mathcal T_{d f h g i j}$ \\ \hline
1344 & $ C_{a b c d} \mathcal T_{a b e f g h} \mathcal T_{c d e i j k} \mathcal T_{f g h i j k}$ \\ \hline
-12096 & $ C_{a b c d} \mathcal T_{a b e f g h} \mathcal T_{c d f i j k} \mathcal T_{e g h i j k} $ \\ \hline
-48384 & $ C_{a b c d} \mathcal T_{a b e f g h} \mathcal T_{c d f i j k} \mathcal T_{e g i h j k}$ \\ \hline
24192 & $ C_{a b c d} \mathcal T_{a b e f g h} \mathcal T_{c e f i j k} \mathcal T_{d g h i j k} $ \\ \hline
2386 & $ \ \mathcal T_{a b c d e f} \mathcal T_{a b c d g h} \mathcal T_{e g i j k l} \mathcal T_{f i j h k l}$ \\ \hline
-3669 & $ \ \mathcal T_{a b c d e f} \mathcal T_{a b c d g h} \mathcal T_{e i j g k l} \mathcal T_{f i k h j l} $ \\ \hline
-1296 & $ \ \mathcal T_{a b c d e f} \mathcal T_{a b c g h i} \mathcal T_{d e j g k l} \mathcal T_{f h k i j l} $ \\ \hline
10368 & $ \ \mathcal T_{a b c d e f} \mathcal T_{a b c g h i} \mathcal T_{d g j e k l} \mathcal T_{f h k i j l} $ \\ \hline
2688 & $ \ \mathcal T_{a b c d e f} \mathcal T_{a b d e g h} \mathcal T_{c g i j k l} \mathcal T_{f j k h i l}$ \\ \hline 
\end{tabular}
\caption{\label{FinalRes}Table detailing the ${\alpha^\prime}^3$ corrections of any solution in type IIB supergravity with nontrivial metric $g$ and five-form $F_{(5)}$. Following \cite{Paulos:2008tn}, all tensor monomials are written with all indices lower.}
\end{table}

We notice that table~\ref{FinalRes} corrects some typos in the final table of \cite{Paulos:2008tn}.

Our objective is to use these results to compute the leading correction to the entropy of the black hole solution detailed in (\ref{eq:5D}). Naively, one might think that we would need to solve the equations of motion from the action (\ref{eq:IIB}) and only then evaluate the correction to the entropy. However, due to the work in \cite{Reall:2019sah} (whose results straightforwardly generalise to the case at hand), one in fact only needs to know the $0^{\text{th}}$ order solution, and evaluate that on the corrected action to get the leading corrections to the entropy.

This is a major simplification and is one of the main reasons this work is possible. However, it is still not a trivial task to evaluate all the monomials from table \ref{FinalRes} without accidentally inserting typos. Therefore, one of the key steps we had to take was validating our calculations. We wrote two pieces of code independently from one another, only comparing them at the end to make sure they agreed. We started by confirming the results of \cite{Paulos:2008tn} to make sure there were no mistakes when copying the monomials from table \ref{FinalRes}.

Only after we had two matching codes that confirmed the results in \cite{Paulos:2008tn} did we insert the solution (\ref{eq:5D}). And even then, to be completely certain we had no typos or no convention compatibility issues, not only did we include many consistency checks throughout the code, \emph{e.g.} confirming we indeed solved the correct equations of motion, but we used two different parametrisations. One of them using a $\mathbb{CP}^2$ fibration and another using a more direct method using the coordinates as originally written in \cite{Cvetic:1999xp}. The $\mathbb{CP}^2$ fibration is the more efficient method and therefore is the one included in the supplemental material. However, the direct method is more amenable to generalisation for the case of different angular momenta \cite{Cvetic:1999xp,Wu:2011gq}, which we leave for future work.

After the colossal amount of dust settles, all twenty terms in table~\ref{FinalRes} are non-vanishing on our solutions, and yet the final result appears simple, which gives further confidence in our answer. Using the relation between the Gibbs free energy $G$ and the Euclidean action obtained from (\ref{eq:IIB}), we find that the stringy correction to the Gibbs free energy at fixed chemical potential $\mu$, angular velocity $\Omega$ and temperature $T$ reads
\begin{multline}
(\delta G)_{\mu,\Omega,T} =-\frac{12 \pi ^3 {\alpha^\prime}^3 \left(\tilde{M}+\tilde{Q}\right)^2 \zeta (3)}{N^2 L^{12} r_+^{15} \left(9 L^2-\tilde{J}^2\right)} \times
\\
 \times \left(L^2-\tilde{J}^2\right)^3\Delta  \left(\Delta +\frac{4}{L}J\right)\leq0\,.
   \label{eq:main}
\end{multline}
It is a simple matter to compute the variation in entropy, $(\delta S)_{Q,J,M}$, at fixed asymptotic charges $Q$, $J$ and $M$ from $(\delta G)_{\mu,\Omega,T}$. In particular, we can follow the same steps as in \cite{Reall:2019sah} to show that
\begin{equation}
    (\delta S)_{Q,J,M} = - T^{-1}(\delta G)_{\mu,\Omega,T}\,.
    \label{eq:susy}
\end{equation}
Equations (\ref{eq:main}) and (\ref{eq:susy}) are the main result of this manuscript, whose physical significance we discuss next.
\\
\noindent\emph{\bf Interpretation of results.}
The first thing we note is the fact that $(\delta S)_{Q,J,M} =0$ on the supersymmetric black hole solutions found in \cite{Gutowski:2004ez}. One might wonder why that is the case, given that (\ref{eq:susy}) has a factor of $T$ in the denominator, and for supersymmetric solutions $T=0$. However, we note that if we take $\tilde{Q}=\tilde{Q}_{\mathrm{BPS}}+ \delta Q$ and $\tilde{J}=\tilde{J}_{\mathrm{BPS}}+ \delta J$, with $\delta Q, \delta J \ll1$, we get $T= \mathcal{O}(\delta Q, \delta J)$, whereas $\Delta =\mathcal{O}(\delta Q^2, \delta Q \delta J, \delta J^2)$. This means $(\delta S)_{Q,J,M}=\mathcal{O}(\delta Q, \delta J)$ in Eq.~(\ref{eq:susy}), \emph{i.e.} it vanishes in the supersymmetric limit. Another way to see this result is to note that one can read off the change in entropy due to stringy corrections at constant chemical potential $\mu$, temperature $T$ and angular velocity $\Omega$ using the standard thermodynamic relation $S = -(\partial G/\partial T)_{\Omega,\mu}$. In this limit, we get that the correction to the entropy is finite at extremality, being zero in the supersymmetric limit. To our knowledge there is no \emph{a priori} reason, based on bulk physics, for why the entropy in the supersymmetic limit is not corrected via stringy effects. This lends support in favour of the index picture advocated in \cite{Cabo-Bizet:2018ehj,Choi:2018hmj,Benini:2018ywd,ArabiArdehali:2019tdm,Zaffaroni:2019dhb,Kim:2019yrz,Cabo-Bizet:2019osg,Lezcano:2019pae,Lanir:2019abx,Cabo-Bizet:2019eaf,Cabo-Bizet:2020nkr,Murthy:2020rbd,Agarwal:2020zwm,Benini:2020gjh,Gadde:2020yah}.

Second, the sign of $(\delta S)_{Q,J,M}$ appears consistent with the weak gravity conjecture \cite{ArkaniHamed:2006dz}, similarly to the analogous calculations in flat space \cite{Kats:2006xp,Hamada:2018dde,Cheung:2018cwt,Loges:2019jzs,Goon:2019faz} and with AdS asymptotics \cite{Cremonini:2019wdk}. In particular, one can show using the generalisation of the Goon-Penco relation to AdS \cite{Goon:2019faz,Cremonini:2019wdk} that the leading correction to the extremality bound at fixed energy $M$, charge $Q$ and angular momentum $J$ necessarily decreases with respect to the uncorrected solution. This relation is in perfect agreement with the weak gravity conjecture \cite{Kats:2006xp,Cheung:2018cwt,Goon:2019faz}.

Thirdly, we point out that our final expression (\ref{eq:main}) only assumes equal angular momenta and equal charges. Notably, it is non-vanishing for a generic non-supersymmetric extremal black hole, and is even valid away from extremality. It would be interesting to understand whether the methods used in \cite{David:2020ems} could be extended to capture the leading $\alpha^\prime$ corrections presented in this letter. Further, this then offers a prediction for the quantum field theoretic calculation. Even though the counting of the supersymmetric states is not corrected at finite $\lambda$, the counting including non-supersymmetric states should be, and its form should be given by (\ref{eq:main}). However, as of yet, there are no techniques capable of computing a partition function at strong coupling without the aid of supersymmetry. Though we should mention that in \cite{Larsen:2019oll} some progress has been reported in going slightly beyond the supersymmetric limit.

Our results rely heavily on \cite{Reall:2019sah}, since we solely use the uncorrected solution to determine the thermodynamic properties of the corrected solution. In principle, we could use the equations of motion that follow from (\ref{eq:IIB}) together with the modified self-duality condition of \cite{Paulos:2008tn} to determine directly the stringy corrected black holes. Under such circumstances, we could determine all thermodynamic properties from the solutions \emph{per se} instead of using the arguments presented in \cite{Reall:2019sah}. Perhaps our current results suggest that the uncorrected supersymmetric solution might be a solution of the corrected equations of motion. This phenomenon has been recently observed in \cite{Bobev:2020egg} for a number of corrections and black hole solutions. We leave this avenue of research for the future.

Finally, an interesting avenue for future work is to generalise this calculation to the case when all the angular momenta and charges are distinct, using the results from \cite{Wu:2011gq}. The complexity of this solution is quite daunting, and computing these corrections would necessarily require more computing power and a more efficient algorithm\footnote{For the interested reader, even just checking that the solution \cite{Wu:2011gq} indeed solves the equations of motion as claimed takes a few hours with a rather optimised \emph{Mathematica} code.}.

\begin{acknowledgments}
\begin{center}
\emph{\bf  Acknowledgments}
\end{center}
We thank Joonho~Kim, Juan~Maldacena, Prahar~Mitra and Edward~Witten for helpful discussions and \'Oscar~Dias, Joonho~Kim, Harvey Reall, Philip Clarke and Maeve Madigan for comments on a draft of this manuscript. J.~E.~S. is supported in part by STFC grants PHY-1504541 and ST/P000681/1.  J.~E.~S. also acknowledges support from a J. Robert Oppenheimer Visiting Professorship. J.~F.~M. thanks the Cambridge Trust for his Vice-Chancellor’s award to support his studies. J.~F.~M. would particularly like to thank many of his friends and family for enduring the frustration with all the 2075 versions of the code that were off by a minus sign or a factor of 2. 
\end{acknowledgments}
\bibliography{refs}{}

\providecommand{\href}[2]{#2}\begingroup\raggedright\begin{thebibliography}{10}

\bibitem{Christodoulou:1970wf}
D.~Christodoulou, {\it {Reversible and irreversible transforations in black
  hole physics}},  {\em Phys. Rev. Lett.} {\bf 25} (1970) 1596--1597.

\bibitem{Hawking:1971tu}
S.~Hawking, {\it {Gravitational radiation from colliding black holes}},  {\em
  Phys. Rev. Lett.} {\bf 26} (1971) 1344--1346.

\bibitem{carter}
B.~Carter, {\it Rigidity of a black hole},  {\em Nature Physical Science} {\bf
  238} (1972), no.~83 71--72.

\bibitem{Bekenstein:1972tm}
J.~Bekenstein, {\it {Black holes and the second law}},  {\em Lett. Nuovo Cim.}
  {\bf 4} (1972) 737--740.

\bibitem{Christodoulou:1972kt}
D.~Christodoulou and R.~Ruffini, {\it {Reversible transformations of a charged
  black hole}},  {\em Phys. Rev. D} {\bf 4} (1971) 3552--3555.

\bibitem{Bekenstein:1973ur}
J.~D. Bekenstein, {\it {Black holes and entropy}},  {\em Phys. Rev. D} {\bf 7}
  (1973) 2333--2346.

\bibitem{Bardeen:1973gs}
J.~M. Bardeen, B.~Carter, and S.~Hawking, {\it {The Four laws of black hole
  mechanics}},  {\em Commun. Math. Phys.} {\bf 31} (1973) 161--170.

\bibitem{Hawking:1974rv}
S.~Hawking, {\it {Black hole explosions}},  {\em Nature} {\bf 248} (1974)
  30--31.

\bibitem{Hawking:1974sw}
S.~Hawking, {\it {Particle Creation by Black Holes}},  {\em Commun. Math.
  Phys.} {\bf 43} (1975) 199--220. [Erratum: Commun.Math.Phys. 46, 206 (1976)].

\bibitem{Strominger:1996sh}
A.~Strominger and C.~Vafa, {\it {Microscopic origin of the Bekenstein-Hawking
  entropy}},  {\em Phys. Lett. B} {\bf 379} (1996) 99--104,
  [\href{http://xxx.lanl.gov/abs/hep-th/9601029}{{\tt hep-th/9601029}}].

\bibitem{Breckenridge:1996sn}
J.~Breckenridge, D.~Lowe, R.~C. Myers, A.~Peet, A.~Strominger, and C.~Vafa,
  {\it {Macroscopic and microscopic entropy of near extremal spinning black
  holes}},  {\em Phys. Lett. B} {\bf 381} (1996) 423--426,
  [\href{http://xxx.lanl.gov/abs/hep-th/9603078}{{\tt hep-th/9603078}}].

\bibitem{Callan:1996dv}
C.~G. Callan and J.~M. Maldacena, {\it {D-brane approach to black hole quantum
  mechanics}},  {\em Nucl. Phys. B} {\bf 472} (1996) 591--610,
  [\href{http://xxx.lanl.gov/abs/hep-th/9602043}{{\tt hep-th/9602043}}].

\bibitem{Horowitz:1996fn}
G.~T. Horowitz and A.~Strominger, {\it {Counting states of near extremal black
  holes}},  {\em Phys. Rev. Lett.} {\bf 77} (1996) 2368--2371,
  [\href{http://xxx.lanl.gov/abs/hep-th/9602051}{{\tt hep-th/9602051}}].

\bibitem{Cyrier:2004hj}
M.~Cyrier, M.~Guica, D.~Mateos, and A.~Strominger, {\it {Microscopic entropy of
  the black ring}},  {\em Phys. Rev. Lett.} {\bf 94} (2005) 191601,
  [\href{http://xxx.lanl.gov/abs/hep-th/0411187}{{\tt hep-th/0411187}}].

\bibitem{Maldacena:1997re}
J.~M. Maldacena, {\it {The Large N limit of superconformal field theories and
  supergravity}},  {\em Int.J.Theor.Phys.} {\bf 38} (1999) 1113--1133,
  [\href{http://xxx.lanl.gov/abs/hep-th/9711200}{{\tt hep-th/9711200}}].

\bibitem{Gubser:1998bc}
S.~S. Gubser, I.~R. Klebanov, and A.~M. Polyakov, {\it {Gauge theory
  correlators from noncritical string theory}},  {\em Phys. Lett.} {\bf B428}
  (1998) 105--114, [\href{http://xxx.lanl.gov/abs/hep-th/9802109}{{\tt
  hep-th/9802109}}].

\bibitem{Witten:1998qj}
E.~Witten, {\it {Anti-de Sitter space and holography}},  {\em Adv. Theor. Math.
  Phys.} {\bf 2} (1998) 253--291,
  [\href{http://xxx.lanl.gov/abs/hep-th/9802150}{{\tt hep-th/9802150}}].

\bibitem{Aharony:1999ti}
O.~Aharony, S.~S. Gubser, J.~M. Maldacena, H.~Ooguri, and Y.~Oz, {\it {Large N
  field theories, string theory and gravity}},  {\em Phys. Rept.} {\bf 323}
  (2000) 183--386, [\href{http://xxx.lanl.gov/abs/hep-th/9905111}{{\tt
  hep-th/9905111}}].

\bibitem{Kinney:2005ej}
J.~Kinney, J.~M. Maldacena, S.~Minwalla, and S.~Raju, {\it {An Index for 4
  dimensional super conformal theories}},  {\em Commun. Math. Phys.} {\bf 275}
  (2007) 209--254, [\href{http://xxx.lanl.gov/abs/hep-th/0510251}{{\tt
  hep-th/0510251}}].

\bibitem{Hosseini:2017mds}
S.~M. Hosseini, K.~Hristov, and A.~Zaffaroni, {\it {An extremization principle
  for the entropy of rotating BPS black holes in AdS$_{5}$}},  {\em JHEP} {\bf
  07} (2017) 106, [\href{http://xxx.lanl.gov/abs/1705.0538}{{\tt
  arXiv:1705.0538}}].

\bibitem{Cabo-Bizet:2018ehj}
A.~Cabo-Bizet, D.~Cassani, D.~Martelli, and S.~Murthy, {\it {Microscopic origin
  of the Bekenstein-Hawking entropy of supersymmetric AdS$_{5}$ black holes}},
  {\em JHEP} {\bf 10} (2019) 062,
  [\href{http://xxx.lanl.gov/abs/1810.1144}{{\tt arXiv:1810.1144}}].

\bibitem{Choi:2018hmj}
S.~Choi, J.~Kim, S.~Kim, and J.~Nahmgoong, {\it {Large AdS black holes from
  QFT}},  \href{http://xxx.lanl.gov/abs/1810.1206}{{\tt arXiv:1810.1206}}.

\bibitem{Benini:2018ywd}
F.~Benini and P.~Milan, {\it {Black holes in 4d $\mathcal{N}=4$
  Super-Yang-Mills}},  {\em Phys. Rev. X} {\bf 10} (2020), no.~2 021037,
  [\href{http://xxx.lanl.gov/abs/1812.0961}{{\tt arXiv:1812.0961}}].

\bibitem{Honda:2019cio}
M.~Honda, {\it {Quantum Black Hole Entropy from 4d Supersymmetric Cardy
  formula}},  {\em Phys. Rev. D} {\bf 100} (2019), no.~2 026008,
  [\href{http://xxx.lanl.gov/abs/1901.0809}{{\tt arXiv:1901.0809}}].

\bibitem{ArabiArdehali:2019tdm}
A.~Arabi~Ardehali, {\it {Cardy-like asymptotics of the 4d $ \mathcal{N}=4 $
  index and AdS$_{5}$ blackholes}},  {\em JHEP} {\bf 06} (2019) 134,
  [\href{http://xxx.lanl.gov/abs/1902.0661}{{\tt arXiv:1902.0661}}].

\bibitem{Zaffaroni:2019dhb}
A.~Zaffaroni, {\it {Lectures on AdS Black Holes, Holography and Localization}},
   in {\em {CERN Winter School on Strings and Fields 2017 and ICTP School on
  Supersymmetric Localization, Holography and Related Topics}}, 2019.
\newblock \href{http://xxx.lanl.gov/abs/1902.0717}{{\tt arXiv:1902.0717}}.

\bibitem{Kim:2019yrz}
J.~Kim, S.~Kim, and J.~Song, {\it {A 4d $N=1$ Cardy Formula}},
  \href{http://xxx.lanl.gov/abs/1904.0345}{{\tt arXiv:1904.0345}}.

\bibitem{Cabo-Bizet:2019osg}
A.~Cabo-Bizet, D.~Cassani, D.~Martelli, and S.~Murthy, {\it {The asymptotic
  growth of states of the 4d $ \mathcal{N}=1 $ superconformal index}},  {\em
  JHEP} {\bf 08} (2019) 120, [\href{http://xxx.lanl.gov/abs/1904.0586}{{\tt
  arXiv:1904.0586}}].

\bibitem{Lezcano:2019pae}
A.~González~Lezcano and L.~A. Pando~Zayas, {\it {Microstate counting via Bethe
  Ansätze in the 4d $ \mathcal{N} $ = 1 superconformal index}},  {\em JHEP}
  {\bf 03} (2020) 088, [\href{http://xxx.lanl.gov/abs/1907.1284}{{\tt
  arXiv:1907.1284}}].

\bibitem{Lanir:2019abx}
A.~Lanir, A.~Nedelin, and O.~Sela, {\it {Black hole entropy function for toric
  theories via Bethe Ansatz}},  {\em JHEP} {\bf 04} (2020) 091,
  [\href{http://xxx.lanl.gov/abs/1908.0173}{{\tt arXiv:1908.0173}}].

\bibitem{Cabo-Bizet:2019eaf}
A.~Cabo-Bizet and S.~Murthy, {\it {Supersymmetric phases of 4d N=4 SYM at large
  N}},  \href{http://xxx.lanl.gov/abs/1909.0959}{{\tt arXiv:1909.0959}}.

\bibitem{Cabo-Bizet:2020nkr}
A.~Cabo-Bizet, D.~Cassani, D.~Martelli, and S.~Murthy, {\it {The large-$N$
  limit of the 4d $\mathcal{N}=1$ superconformal index}},
  \href{http://xxx.lanl.gov/abs/2005.1065}{{\tt arXiv:2005.1065}}.

\bibitem{Murthy:2020rbd}
S.~Murthy, {\it {The growth of the $\frac{1}{16}$-BPS index in 4d
  $\mathcal{N}=4$ SYM}},  \href{http://xxx.lanl.gov/abs/2005.1084}{{\tt
  arXiv:2005.1084}}.

\bibitem{Agarwal:2020zwm}
P.~Agarwal, S.~Choi, J.~Kim, S.~Kim, and J.~Nahmgoong, {\it {AdS black holes
  and finite N indices}},  \href{http://xxx.lanl.gov/abs/2005.1124}{{\tt
  arXiv:2005.1124}}.

\bibitem{Benini:2020gjh}
F.~Benini, E.~Colombo, S.~Soltani, A.~Zaffaroni, and Z.~Zhang, {\it
  {Superconformal indices at large $N$ and the entropy of AdS$_5$ $\times$
  SE$_5$ black holes}},  \href{http://xxx.lanl.gov/abs/2005.1230}{{\tt
  arXiv:2005.1230}}.

\bibitem{Gadde:2020yah}
A.~Gadde, {\it {Lectures on the Superconformal Index}},
  \href{http://xxx.lanl.gov/abs/2006.1363}{{\tt arXiv:2006.1363}}.

\bibitem{Gutowski:2004ez}
J.~B. Gutowski and H.~S. Reall, {\it {Supersymmetric AdS(5) black holes}},
  {\em JHEP} {\bf 02} (2004) 006,
  [\href{http://xxx.lanl.gov/abs/hep-th/0401042}{{\tt hep-th/0401042}}].

\bibitem{Gutowski:2004yv}
J.~B. Gutowski and H.~S. Reall, {\it {General supersymmetric AdS(5) black
  holes}},  {\em JHEP} {\bf 04} (2004) 048,
  [\href{http://xxx.lanl.gov/abs/hep-th/0401129}{{\tt hep-th/0401129}}].

\bibitem{Kunduri:2006ek}
H.~K. Kunduri, J.~Lucietti, and H.~S. Reall, {\it {Supersymmetric multi-charge
  AdS(5) black holes}},  {\em JHEP} {\bf 04} (2006) 036,
  [\href{http://xxx.lanl.gov/abs/hep-th/0601156}{{\tt hep-th/0601156}}].

\bibitem{Bates:2003vx}
B.~Bates and F.~Denef, {\it {Exact solutions for supersymmetric stationary
  black hole composites}},  {\em JHEP} {\bf 11} (2011) 127,
  [\href{http://xxx.lanl.gov/abs/hep-th/0304094}{{\tt hep-th/0304094}}].

\bibitem{Dabholkar:2012nd}
A.~Dabholkar, S.~Murthy, and D.~Zagier, {\it {Quantum Black Holes, Wall
  Crossing, and Mock Modular Forms}},
  \href{http://xxx.lanl.gov/abs/1208.4074}{{\tt arXiv:1208.4074}}.

\bibitem{Paulos:2008tn}
M.~F. Paulos, {\it {Higher derivative terms including the Ramond-Ramond
  five-form}},  {\em JHEP} {\bf 10} (2008) 047,
  [\href{http://xxx.lanl.gov/abs/0804.0763}{{\tt arXiv:0804.0763}}].

\bibitem{Green:2003an}
M.~B. Green and C.~Stahn, {\it {D3-branes on the Coulomb branch and
  instantons}},  {\em JHEP} {\bf 09} (2003) 052,
  [\href{http://xxx.lanl.gov/abs/hep-th/0308061}{{\tt hep-th/0308061}}].

\bibitem{Cvetic:2004hs}
M.~Cvetic, H.~Lu, and C.~Pope, {\it {Charged Kerr-de Sitter black holes in five
  dimensions}},  {\em Phys. Lett. B} {\bf 598} (2004) 273--278,
  [\href{http://xxx.lanl.gov/abs/hep-th/0406196}{{\tt hep-th/0406196}}].

\bibitem{York:1972sj}
J.~York, James~W., {\it {Role of conformal three geometry in the dynamics of
  gravitation}},  {\em Phys. Rev. Lett.} {\bf 28} (1972) 1082--1085.

\bibitem{Gibbons:1976ue}
G.~Gibbons and S.~Hawking, {\it {Action Integrals and Partition Functions in
  Quantum Gravity}},  {\em Phys. Rev. D} {\bf 15} (1977) 2752--2756.

\bibitem{Balasubramanian:1999re}
V.~Balasubramanian and P.~Kraus, {\it {A Stress tensor for Anti-de Sitter
  gravity}},  {\em Commun. Math. Phys.} {\bf 208} (1999) 413--428,
  [\href{http://xxx.lanl.gov/abs/hep-th/9902121}{{\tt hep-th/9902121}}].

\bibitem{deHaro:2000vlm}
S.~de~Haro, S.~N. Solodukhin, and K.~Skenderis, {\it {Holographic
  reconstruction of space-time and renormalization in the AdS / CFT
  correspondence}},  {\em Commun. Math. Phys.} {\bf 217} (2001) 595--622,
  [\href{http://xxx.lanl.gov/abs/hep-th/0002230}{{\tt hep-th/0002230}}].

\bibitem{Loges:2020trf}
G.~J. Loges, T.~Noumi, and G.~Shiu, {\it {Duality and Supersymmetry Constraints
  on the Weak Gravity Conjecture}},
  \href{http://xxx.lanl.gov/abs/2006.0669}{{\tt arXiv:2006.0669}}.

\bibitem{Markeviciute:2016ivy}
J.~Markeviciute and J.~E. Santos, {\it {Hairy black holes in
  AdS$_{5}\times$S$^{5}$}},  {\em JHEP} {\bf 06} (2016) 096,
  [\href{http://xxx.lanl.gov/abs/1602.0389}{{\tt arXiv:1602.0389}}].

\bibitem{Markeviciute:2018cqs}
J.~Markeviciute, {\it {Rotating Hairy Black Holes in AdS$_5\times$S$^5$}},
  {\em JHEP} {\bf 03} (2019) 110,
  [\href{http://xxx.lanl.gov/abs/1809.0408}{{\tt arXiv:1809.0408}}].

\bibitem{Chamblin:1999tk}
A.~Chamblin, R.~Emparan, C.~V. Johnson, and R.~C. Myers, {\it {Charged AdS
  black holes and catastrophic holography}},  {\em Phys. Rev. D} {\bf 60}
  (1999) 064018, [\href{http://xxx.lanl.gov/abs/hep-th/9902170}{{\tt
  hep-th/9902170}}].

\bibitem{Cvetic:1999xp}
M.~Cvetic, M.~Duff, P.~Hoxha, J.~T. Liu, H.~Lu, J.~Lu, R.~Martinez-Acosta,
  C.~Pope, H.~Sati, and T.~A. Tran, {\it {Embedding AdS black holes in
  ten-dimensions and eleven-dimensions}},  {\em Nucl. Phys. B} {\bf 558} (1999)
  96--126, [\href{http://xxx.lanl.gov/abs/hep-th/9903214}{{\tt
  hep-th/9903214}}].

\bibitem{Cvetic:2000nc}
M.~Cvetic, H.~Lu, C.~Pope, A.~Sadrzadeh, and T.~A. Tran, {\it {Consistent SO(6)
  reduction of type IIB supergravity on S**5}},  {\em Nucl. Phys. B} {\bf 586}
  (2000) 275--286, [\href{http://xxx.lanl.gov/abs/hep-th/0003103}{{\tt
  hep-th/0003103}}].

\bibitem{Reall:2019sah}
H.~S. Reall and J.~E. Santos, {\it {Higher derivative corrections to Kerr black
  hole thermodynamics}},  {\em JHEP} {\bf 04} (2019) 021,
  [\href{http://xxx.lanl.gov/abs/1901.1153}{{\tt arXiv:1901.1153}}].

\bibitem{Wu:2011gq}
S.-Q. Wu, {\it {General Nonextremal Rotating Charged AdS Black Holes in
  Five-dimensional $U(1)^3$ Gauged Supergravity: A Simple Construction
  Method}},  {\em Phys. Lett. B} {\bf 707} (2012) 286--291,
  [\href{http://xxx.lanl.gov/abs/1108.4159}{{\tt arXiv:1108.4159}}].

\bibitem{ArkaniHamed:2006dz}
N.~Arkani-Hamed, L.~Motl, A.~Nicolis, and C.~Vafa, {\it {The String landscape,
  black holes and gravity as the weakest force}},  {\em JHEP} {\bf 06} (2007)
  060, [\href{http://xxx.lanl.gov/abs/hep-th/0601001}{{\tt hep-th/0601001}}].

\bibitem{Kats:2006xp}
Y.~Kats, L.~Motl, and M.~Padi, {\it {Higher-order corrections to mass-charge
  relation of extremal black holes}},  {\em JHEP} {\bf 12} (2007) 068,
  [\href{http://xxx.lanl.gov/abs/hep-th/0606100}{{\tt hep-th/0606100}}].

\bibitem{Hamada:2018dde}
Y.~Hamada, T.~Noumi, and G.~Shiu, {\it {Weak Gravity Conjecture from Unitarity
  and Causality}},  {\em Phys. Rev. Lett.} {\bf 123} (2019), no.~5 051601,
  [\href{http://xxx.lanl.gov/abs/1810.0363}{{\tt arXiv:1810.0363}}].

\bibitem{Cheung:2018cwt}
C.~Cheung, J.~Liu, and G.~N. Remmen, {\it {Proof of the Weak Gravity Conjecture
  from Black Hole Entropy}},  {\em JHEP} {\bf 10} (2018) 004,
  [\href{http://xxx.lanl.gov/abs/1801.0854}{{\tt arXiv:1801.0854}}].

\bibitem{Loges:2019jzs}
G.~J. Loges, T.~Noumi, and G.~Shiu, {\it {Thermodynamics of 4D Dilatonic Black
  Holes and the Weak Gravity Conjecture}},  {\em Phys. Rev. D} {\bf 102} (2020)
  046010, [\href{http://xxx.lanl.gov/abs/1909.0135}{{\tt arXiv:1909.0135}}].

\bibitem{Goon:2019faz}
G.~Goon and R.~Penco, {\it {Universal Relation between Corrections to Entropy
  and Extremality}},  {\em Phys. Rev. Lett.} {\bf 124} (2020), no.~10 101103,
  [\href{http://xxx.lanl.gov/abs/1909.0525}{{\tt arXiv:1909.0525}}].

\bibitem{Cremonini:2019wdk}
S.~Cremonini, C.~R. Jones, J.~T. Liu, and B.~McPeak, {\it {Higher-Derivative
  Corrections to Entropy and the Weak Gravity Conjecture in Anti-de Sitter
  Space}},  \href{http://xxx.lanl.gov/abs/1912.1116}{{\tt arXiv:1912.1116}}.

\bibitem{David:2020ems}
M.~David, J.~Nian, and L.~A. Pando~Zayas, {\it {Gravitational Cardy Limit and
  AdS Black Hole Entropy}},  \href{http://xxx.lanl.gov/abs/2005.1025}{{\tt
  arXiv:2005.1025}}.

\bibitem{Larsen:2019oll}
F.~Larsen, J.~Nian, and Y.~Zeng, {\it {AdS$_{5}$ black hole entropy near the
  BPS limit}},  {\em JHEP} {\bf 06} (2020) 001,
  [\href{http://xxx.lanl.gov/abs/1907.0250}{{\tt arXiv:1907.0250}}].

\bibitem{Bobev:2020egg}
N.~Bobev, A.~M. Charles, K.~Hristov, and V.~Reys, {\it {The Unreasonable
  Effectiveness of Higher-Derivative Supergravity in AdS$_4$ Holography}},
  \href{http://xxx.lanl.gov/abs/2006.0939}{{\tt arXiv:2006.0939}}.

\end{thebibliography}\endgroup
\bibliographystyle{JHEP}

\end{document}